\documentclass[12pt]{article}
\usepackage{amsmath,latexsym,epsf}
\usepackage{epsfig,amsfonts,cite}
\usepackage{amssymb}
\usepackage{graphicx}

\begin{document}

\title{Topological transition in a
	two-dimensional model of liquid crystal}
\date{\today}
\author{
  {Ana I. Fari\~nas-S\'anchez$^{*,**}$, Ricardo Paredes$^{*}$ 
	and  Bertrand Berche$^{*,**}$}\\
  {\small\it $^*$ Centro de F\'\i sica,}\\[-0.2cm]
  {\small\it Instituto Venezolano de Investigaciones 
	Cient\'\i ficas,}\\[-0.2cm]
  {\small\it Apartado 21827, Caracas 1020A, Venezuela} 	\\[-0.2cm]
  \\
  {\small\it $^{**}$ Laboratoire de Physique des Mat\'eriaux, }  \\[-0.2cm]
  {\small\it Universit\'e Henri Poincar\'e, Nancy 1,}     \\[-0.2cm]
  {\small\it BP 239, F-54506 Vand\oe uvre les Nancy Cedex, France} \\[-0.2cm]
  \\
  {\small {\tt afarinas@ivic.ve }}\\[-0.2cm]
  {\small {\tt rparedes@ivic.ve}}\\[-0.2cm]
  {\small {\tt berche@lpm.u-nancy.fr }}\\[-0.2cm]
  {\protect\makebox[5in]{\quad}}  
  \\
}
\vspace{0.5cm}

\newcommand\bsigma{\mbox{\boldmath $\sigma$}}
\newcommand\bmu{\mbox{\boldmath $\mu$}}
\newcommand\bsmu{{\mbox{\scriptsize\boldmath $\mu$}}}
\newcommand{\be}{\begin{equation}}
\newcommand{\ee}{\end{equation}}
\newcommand{\<}{\langle}
\renewcommand{\>}{\rangle}
\renewcommand{\r}{{\bf r}}

\maketitle
\thispagestyle{empty}   

\vspace{0.2cm}

\begin{abstract}
Simulations of nematic-isotropic transition of liquid crystals
in two dimensions are performed using an $O(2)$ vector model 
characterised by non linear nearest neighbour spin 
interaction governed by the fourth Legendre polynomial $P_4$.
The system is studied through standard Finite-Size Scaling
and conformal rescaling of density profiles of correlation functions. 
A topological transition between a paramagnetic phase at high temperature
and a critical phase at low temperature is observed.
The low temperature limit is 
discussed in the spin wave approximation and confirms
the numerical results.

\end{abstract}

\section{Introduction\label{Sec:Intro}}
The molecules of liquid crystals may often be described by long, neutral 
rigid rods which interact through 
electrostatic dipolar or higher order multi-polar interactions.
This is at the origin of the natural introduction of Legendre polynomials
for the description of the orientational transition between a disordered,
isotropic high temperature phase and an ordered nematic phase
at lower temperature~\cite{MaierSaupe59}.
When such a material is cooled at even lower temperatures, other
ordered phases may be encountered, the description of which requires more 
realistic potentials (see e.g. Ref.~\cite{Zewdie00}).

Lattice models of nematic-isotropic transitions capture the essentials
of the above description. The molecules are represented by $n-$component unit
vectors $\bsigma_\r$, hereafter called ``spins'', located on the sites $\r$ 
of a simple 
hyper-cubic lattice.
The interaction between molecules is restricted to the nearest neighbours
$\<\r,\r'\>$, so that the radial dependence is kept constant and the angular 
dependence enters through a $k-$th order Legendre polynomial\footnote{Even
order Legendre polynomials guarantee the local $Z_2$ symmetry 
$\bsigma_\r\to -\bsigma_\r$.}, 
$P_k(\cos\alpha_{\r,\r'})$, where $\alpha_{\r,\r'}
=\widehat{(\bsigma_\r,\bsigma_{\r'})}$ is the angle 
between vectors $\bsigma_\r$ and $\bsigma_{\r'}$. The intensity 
of the interaction energy is measured through a parameter $\epsilon$. The
Hamiltonian of a lattice liquid crystal is thus given by
\be
	H=-\epsilon\sum_{\<\r,\r'\>}P_k(\cos\alpha_{\r,\r'}),\label{eq:defH}
\ee
and the relevant parameters for the investigation of the phase transition are
the space dimension $D$, the ``spin'' dimensionality or equivalently, its
symmetry
$O(n)$, and the ``symmetry'' $k$ 
of the interaction $P_k$.

Usual $XY$ and Heisenberg models correspond, within this description,
to $k=1$ and $n=2$ and 3, respectively. In two dimensions ($D=2$), these
models exhibit quite different behaviours. 
In the case of the $XY$ model 
($O(2)$ abelian symmetry), a topological 
transition described by Berezinski\u\i, and Kosterlitz and 
Thouless~\cite{Berezinskii71,KosterlitzThouless73,Kosterlitz74} takes 
place, governed by the condensation of vortices.
This is not prevented by the Mermin-Wagner-Hohenberg 
theorem~\cite{MerminWagner66,Hohenberg67,GelfertNolting01} which states 
that spin models with continuous symmetry cannot have any long range 
ordered phase. To leading order in the high-temperature expansion, one 
gets for the correlation function an 
exponential 
decay $\<\bsigma_{\r_1}\cdot\bsigma_{\r_2}\>\sim K^{|\r_1-\r_2|}$ 
($K=\epsilon/k_BT$), while at low temperatures 
in the harmonic approximation, 
the Hamiltonian becomes quadratic and
leads to the Gaussian
model which implies~\cite{Rice65} that 
$\<\bsigma_{\r_1}\cdot\bsigma_{\r_2}\>\simeq|\r_1-\r_2|^{-1/2\pi K}$, i.e.
a temperature-dependent spin-spin critical exponent
$\eta_{XY}(T)=\frac{k_BT}{2\pi \epsilon},\quad T\to 0$. 
The low temperature (LT) phase of the $XY$ model is a 
quasi-long-range ordered phase (QLRO)  with vanishing magnetisation
$M^2(T)=\lim_{|\r_1-\r_2|\to\infty}\<\bsigma_{\r_1}\cdot\bsigma_{\r_2}\>=0$, 
or a critical phase.
The Heisenberg model ($O(3)$, non-abelian symmetry) on the other hand
has no transition at any finite temperature 
(asymptotic freedom)~\cite{Polyakov75,BrezinZinnJustin76,Amit84,Izyumov88}. 
This difference with $XY$ model is at first surprising, since
the low temperature limit of the Heisenberg model is essentially described
by a similar spin wave approximation (SWA): the longitudinal
modes of the $O(n)$ model are frozen and only the transverse modes are
activated, leading essentially to two Gaussian 
models.
This apparent contradiction between the asymptotic
freedom of the Heisenberg model and the topological transition at finite
temperature for the $XY$
model finds its origins in the stability of topological defects in the latter
case, while the `third spin dimension' makes the vortices unstable at any
temperature in the former model.
Further, the question of the accessibility
of the thermodynamic limit is worth studying. 
Berezinski\u\i\  and Blank noticed
long time ago that a really large, but finite $XY$ system  always 
possesses a {\em non-zero
magnetisation}~\cite{BerezinskiiBlank73,BramwellHoldsworth93}.
The size being limited, $|\r_1-\r_2|$ becomes at most as large as the linear
size $L$ and
a finite order parameter follows, $M_L(T)\sim L^{-\frac 12\eta(T)}$. More
precisely, $M_L(T)=O(1)$ as long as $L\ll {\rm e}^{2/\eta(T)}$ a condition
which can be fulfilled for any $L$ by considering small enough temperatures.
With this result in mind, we expect for the model considered hereafter
that a spin wave solution will be found at low 
enough temperature and thus
we search for an algebraic decay of the spin-spin correlation function from
an effectively {\em non-zero} order parameter, although there is at most 
QLRO at low temperatures.

Changing the value of $k$ in Eq.~(\ref{eq:defH}) 
modifies the symmetry of the spin-spin
interaction and gives rise to new features\footnote{In the same 
spirit, the case of 
symmetry-breaking magnetic fields $h_k\cos k\theta$ added to the $XY$ model and
changing the phase diagram was investigated by Jos\'e et 
al~\cite{JoseEtAl77,Nelson02}.}. 
When $k$ increases, one may  indeed expect a qualitative change 
in the nature of 
the transition, like in the case of discrete spin symmetries 
(Potts model)~\cite{DomanySchickSwendsen84,JonssonMinnhagenNylen93}. 
The value $k=2$ was intensively studied. It still corresponds to the $XY$ model
for $O(2)$ spin symmetry, while it leads to the $RP^2$ or 
Lebwohl-Lasher model~\cite{LebwohlLasher73} for $3-$component spin vectors. 
The nature of the transition in this
latter case is still under discussion~\cite{KunzZumbach92,BloteGuoHilhorst02,CaraccioloPelissetto02,FarinasParedesBerche03,FarinasParedesBerche04}, but
a recent study reported new evidences, extremely convincing, in favour 
of a topological transition~\cite{DuttaRoy04}. The transition is driven
by topologically stable point defects known as $\frac 12$ disclination points.
Considering still larger values of $k$, there is a proof of asymptotic freedom
in the large$-n$ limit,  for values
of $k$ (in the interaction term $(1+\cos\theta)^k$) 
which do not exceed a critical 
$k_c\simeq 4.537...$~\cite{CaraccioloPelissetto02}. This is again 
discussed in a recent preprint from Caracciolo et 
al.~\cite{CaraccioloMognettiPelissetto04}.
Above this value the transition becomes of first order, 
a result which does not violate 
Mermin-Wagner-Hohenberg theorem, since the correlation length is finite
at the transition.
For finite value of $n$, the question of the nature of the transition
at high $k$ is still a challenging problem. In the context of orientational
transitions in liquid crystals, Legendre polynomials rather than $\cos^k\theta$
interactions are introduced, and we are led to the Hamiltonian of 
Eq.~(\ref{eq:defH}). 
The 3-vector model with $P_4$ interactions was already considered
in Refs.~\cite{MukhopadhyayPalRoy99,PalRoy03} where convincing
evidence for a first order transition was reported. 

In this paper,  
we study the behaviour of an abelian
spin model, namely $O(2)$ 
rotation group with
$P_4-$like spin interactions. 
The model will be referred to as $P_4\ O(2)$
for simplicity.
We are mainly interested in the low temperature properties of the model
where comparisons with analytic predictions are possible due to the simplicity
of the SW approximation. The techniques used combine
temperature analysis, 
FSS (Finite-Size Scaling) and 
conformal techniques (Finite-Shape Scaling - FShS - to plagiarize the famous
acronym).

\section{Definition of the model and of the observables}
\label{Sec:Model}
In Refs.~\cite{MukhopadhyayPalRoy99,PalRoy03}, the following Hamiltonian
	$H_{P_4\ O(3)}=
	-\epsilon\sum_{\<\r,\r'\>}P_4(\bsigma_\r\cdot\bsigma_{\r'})$
was considered,
where  
$\bsigma_\r=(\sigma_\r^x,\sigma_\r^y,
\sigma_\r^z)$, $|\bsigma_\r|=1$ and 
$P_4(x)=\frac 18(35x^4-30x^2+3)$.
For $2-$component vectors, in the completely disordered phase 
$\<\cos^2\theta\>=\frac 12$
and $\<\cos^4\theta\>=\frac 38$. In order to keep the same symmetry in
the interaction than in the $P_4\ O(3)$ model, but to 
normalise it between 0 and 1 in the limits of 
completely disordered and completely ordered phases respectively, we 
modify slightly the Hamiltonian to include pair interactions of the type
$Q_4(x)\equiv AP_4(x)+{\rm const}=\frac8{55}(35x^4-30x^2+\frac{15}8)$. The
corresponding Hamiltonian is thus defined by
\be
	H_{P_4\ O(2)}=
	-\epsilon\sum_{\<\r,\r'\>}Q_4(\bsigma_\r\cdot\bsigma_{\r'}),
	\label{eq:defHO2}
\ee
with now 
$\bsigma_\r=(\sigma_\r^x,\sigma_\r^y)$, $|\bsigma_\r|=1$.
A qualitative description of the transition  
is provided by the temperature behaviour of the energy density, the specific
heat, the order parameters  and the corresponding susceptibilities.
The internal energy is defined from the thermal average of the Hamiltonian 
density, 
\be
	u_{P_4\ O(2)}(T)=(DL^D)^{-1}\<H_{P_4\ O(2)}\>
	\label{eq:U}
\ee
and the specific
heat follows from fluctuation dissipation theorem, 
\be
	L^DT^2C_v(T)=\<(H_{P_4\ O(2)})^2\>-\<H_{P_4\ O(2)}\>^2
	\label{eq:Cv}.
\ee 
Brackets denote the thermal average.
The definition of the scalar order parameter (sometimes called nematisation) 
is deduced from the local
second-rank order parameter tensor, 
\be q^{\alpha\beta}(\r)=\sigma_\r^\alpha
\sigma_\r^\beta-\frac 12\delta^{\alpha\beta}.\ee
After space average,
the traceless tensor  $L^{-D}\sum_\r q^{\alpha\beta}(\r)$ 
admits two opposite eigenvalues $\pm\frac 12\eta$ corresponding to eigenvectors
${\bf n }_+$ and ${\bf n}_-$.
The order parameter density is defined after thermal averaging by
\be q_2(T)=\<\eta\>.\label{eq:q2}\ee 
This quantity has the same physical content but is more stable numerically 
than a direct estimation of
$\<2(\bsigma_\r\cdot{\bf n}_+)^2-1\>$.
Another order parameter may
be defined simply by inspection of the structure of the Hamiltonian,
\be q_4(T)=L^{-D}\<\sum_{\r}Q_4(\bsigma_\r\cdot{\bf n}_+)\>.\label{eq:q4}\ee
The associated susceptibilities are defined by the fluctuations of the order
parameter densities, e.g. 
\be\chi_{q_2}(T)=\frac{4L^D}{k_BT}(\<\eta^2\>-\<\eta\>^2).\label{eq:chi}\ee

\section{Thermal behaviour}\label{Sec:Thermal}
In this section, we illustrate the behaviour of the various thermodynamic 
quantities  
as the temperature varies. 
It gives a {\em first idea} of the nature of the transition.

The simulations are performed using a standard Wolff algorithm suited to the
expression of the nearest neighbour interaction~\cite{Wolff89,PalRoy03}. 
The spins are located on the 
vertices of a simple square lattice of size $L^2$ with periodic boundary 
conditions in the two directions.
We use $L=24$, $32$, $48$, $64$ and $128$ with 
$10^6$ equilibrium steps (measured as the number of flipped Wolff clusters)
and $10^6$ Monte Carlo steps (MCS) for the evaluation
of thermal averages. 
The autocorrelation time (at $k_BT/\epsilon=0.2$, $L=16$)
is of order of $30$~MCS, hence the numbers of iterations 
that we used correspond roughly to
$3.10^4$ independent measurements for the smallest size and is still
safe at $L=128$. 

%%% figure %%%
\begin{figure}[h]
  \centering
  \begin{minipage}{\textwidth}
  \epsfig{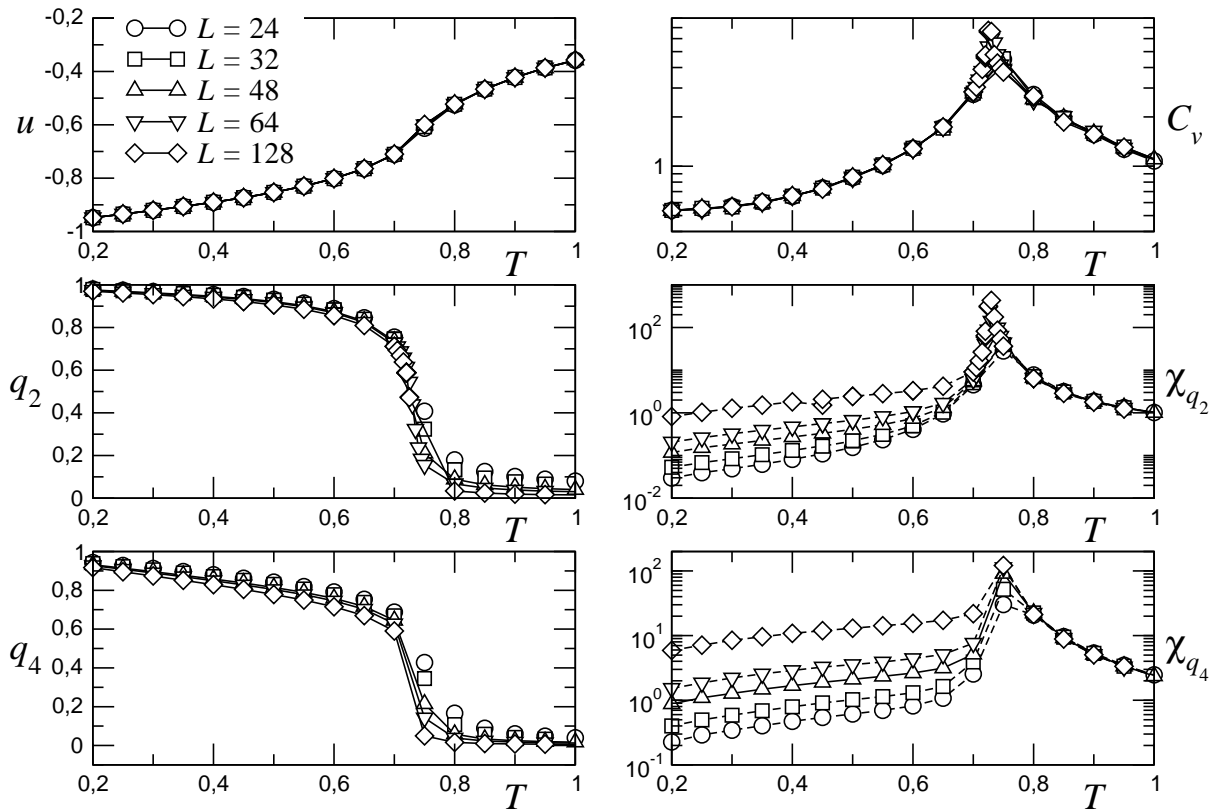}%\qquad
  \end{minipage}
  \caption{Energy, specific heat, order parameter $q_2$ and corresponding 
	susceptibility $\chi_{q_2}$, order parameter $q_4$ and corresponding 
	susceptibility $\chi_{q_4}$ vs $T$ for the $P_4\ O(2)$ model. 
	The full lines are only guides for the eyes. The values
	of $k_B$ and $\epsilon$ have been fixed to unity.}
	\label{Fig1}
\end{figure}
%%% end figure %%%

We deliberately did not use histogram reweighting which
would require a huge amount of simulation time to be 
reliable. We have 
adopted a different strategy here,
producing more simulations with less iterations, but with a 
better control of the errors than at
the ends of the histograms.
The temperature dependence of thermodynamic quantities 
is plotted in Fig.~\ref{Fig1} for different system sizes.
The behaviour of the energy density  clearly displays
a difference between the regimes of low and high temperatures. This is 
the signature of a transition and this naive statement is corroborated by the
behaviours of the other physical quantities, the specific heat $C_v$, the
order parameters $q_2$ and $q_4$, and the corresponding susceptibilities.
The specific heat close to the maximum does not seem to  
increase substantially with the 
system size. This might be the sign of an essential 
singularity~\footnote{Or at least the
sign of a non diverging specific heat with non-positive exponent $\alpha$.}
around a temperature $k_BT_c/\epsilon\simeq 0.70-0.75$.
From the behaviour of the order parameters, 
we may suspect a smooth transition, since there is
no sharp jump. The susceptibilities display a non conventional 
behaviour at low temperature, increasing with the system size, 
which indicates a likely topological
transition with a critical low temperature phase where the susceptibility
diverges at any temperature (note the logarithmic scale for vertical axis).

\begin{figure}[ht]
  \centering
  \begin{minipage}{\textwidth}
  \epsfig{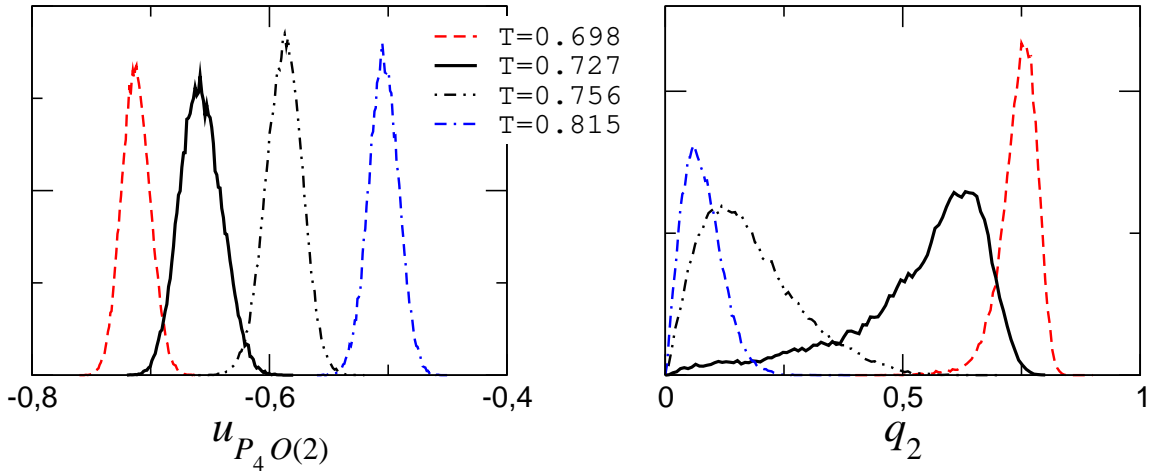}
  \end{minipage}
  \caption{Energy density and order parameter probability distributions
	(arbitrarily 	normalized) for the $P_4\ O(2)$ 
	model
	at a temperature close to the transition .
	The system size is $L=48$ 
	and the temperatures are given in units suche that $k_B/\epsilon=1$. 
	The solid lines 
	correspond to the effective transition.}
	\label{Fig3bis}
\end{figure}

The probability distributions of the energy density and the order parameter
is also instructive. Both quantities are shown 
in Fig.~\ref{Fig3bis}. 
The distributions
have a simple shape with single peaks, and this is still true for temperatures
below the transition, a result which suggests a continuous transition.
We note that due to the finite-size of the system, the order parameter 
is finite. 
%%% figure %%%
%%% end figure %%%

\section{Finite-Size Scaling}\label{Sec:FSS}
In this section we investigate the properties of the low temperature phase
using Finite-Size Scaling technique.
In the critical low temperature phase of a model which displays a topological
transition (the paradigmatic $XY$ model serves as a guide), the physical
quantities behave like at criticality for a second-order phase transition, 
with power law behaviours of the system size. The difference is that in the
critical phase, the critical exponents depend on the temperature and for any
temperature below the transition one
has e.g. 
\be q_2(T)\sim L^{-\frac 12\eta_{q_2}(T)}\label{eq-q2}\ee 
\be\chi_{q_2}(T)\sim L^{2-\eta_{q_2}(T)}.\label{eq-chiq2}\ee 
Here $\eta_{q_2}(T)$ denotes the correlation
function critical exponent, defined by 
\be
\<Q_2(\cos(\theta_{\r_1}-\theta_{\r_2}))\>\sim|\r_1-\r_2|^{-\eta_{q_2}(T)}.
\ee

%%% figure %%%
\begin{figure}[ht]
  \centering
  \begin{minipage}{\textwidth}
  \epsfig{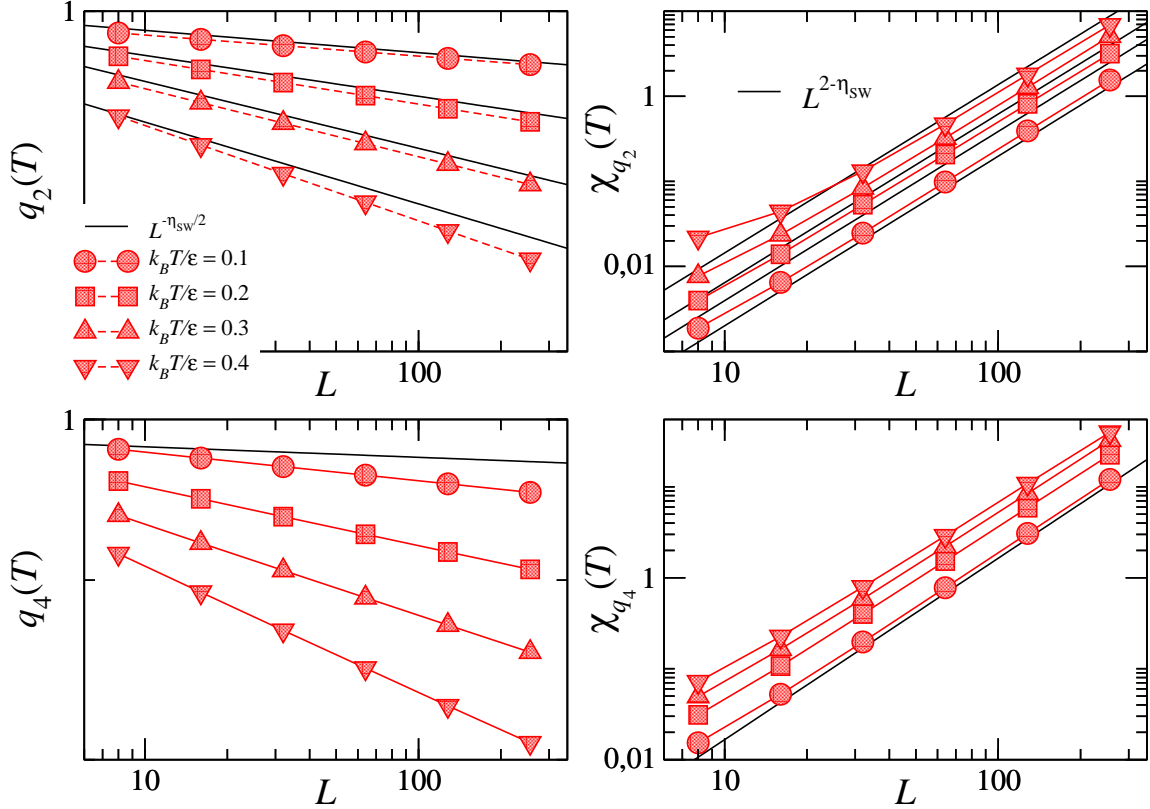}
  \end{minipage}
  \caption{FSS behaviour of the order parameter densities $q_2(T)$ and 
	$q_4(T)$ and the corresponding
	susceptibilities for the $P_4\ O(2)$ model at temperatures
	$k_BT/\epsilon=0.1$, $0.2$ and $0.4$ below the transition temperature.
	Temperature-dependent exponents are confirmed and 
	the full lines show the expected SW result which nicely fits
	the data at low temperatures.}
	\label{Fig3}
\end{figure}
%%% end figure %%%

In Fig.~\ref{Fig3}  
the FSS behaviour of the order parameter
densities and of the corresponding susceptibilities is shown on log-log
scales vs the system size $L$ (we used $L=16$,  32, 48, 64, 128 and 256) for
different values of $T$ below the expected transition.
The linear behaviour on this scale indicates power laws, and the different
slopes at different temperatures 
is the result of exponents which depend
on the value of the temperature.
Starting from a fully ordered system at $T=0$, at low temperature 
we can expect a small disorientation of the molecules
and the spin-wave approximation becomes correct.
For $O(2)$ model with nearest neighbour interactions described by
arbitrary polynomial in $\cos\alpha_{\r,\r'}$, one is led to 
an effective harmonic Hamiltonian $\frac 12\epsilon\sum_{\<\r,\r'\>}l
(\theta_\r-\theta_{\r'})^2$.
It yields power-law correlations,
\be
	\<\cos m(\theta_{\r_1}-\theta_{\r_2})\>
	={\rm e}^{
 	-{\textstyle \frac {m^2}{2}}\langle(\theta_{\r_1}-\theta_{\r_2})^2
	\rangle} 
	\sim|\r_1-\r_2|^{-\eta^{\rm SW}_{ml}}
	\label{eq:SWcorrel}
\ee
with a spin-wave decay exponent given by 
\be
	\eta^{\rm SW}_{ml}=\frac {m^2}{l}\eta_{XY}(T)=
	\frac{m^2k_BT}{2l\pi\epsilon}.
	\label{eq:eta}
\ee 
This expression, if confirmed
numerically, will support the presence of a quasi-long-range 
ordered,
scale-invariant phase at low temperatures.
The exponent 
$\eta_{q_2}(T)$ ($m=2$) is 
accessible through $q_2(T)$ (or the
corresponding susceptibility) via equations~(\ref{eq-q2}) and 
(\ref{eq-chiq2}). 
We note that it is also accessible through the
the FSS behaviour of $q_4(T)$, since the leading behaviour of
$\<Q_4(\theta_{\r_1}-\theta_{\r_2})\>$ is still governed by 
$\<\cos 2(\theta_{\r_1}-\theta_{\r_2})\>$. 
In Fig~\ref{Fig3}, one may extract  numerical
values of $\eta$ exponent. The  
values $\eta_{q_2}(T)=0.0056$, $0.0117$, $0.0182$, and 
$0.0256$ are obtained at $k_BT/\epsilon = 0.1$, $0.2$, $0.3$,  
and $0.4$, respectively.
Using the small angle limit 
$H_{P_4\ O(2)}=-\epsilon\sum_{\<\r,\r'\>} Q_4
	(\cos\alpha_{\r,\r'})\simeq \frac 12\epsilon\sum_{\<\r,\r'\>}
	\frac {128}{11}(\theta_\r-\theta_{\r'})^2,$
we have $l=\frac {128}{11}$ which yields 
$\eta^{\rm SW}_{2\frac{128}{11}}=\frac{11k_BT}{64\pi\epsilon}$. 
At $k_BT/\epsilon=0.1$, $0.2$, $0.3$,
and 0.4, we get
$\eta^{\rm SW}_{2\frac{128}{11}}(T)\simeq 0.006$, $0.011$, $0.018$, 
and $0.022$ which 
confirm the numerical values, since the SW exponent is always a lower 
bound for the exact exponent $\eta_{q_2}(T)$.
The comparison is made visible in the figure where the SW values are
plotted in full lines while the symbols represent the numerical data.
As expected, the lower the temperature, the better the SW approximation.

\section{Finite-Shape Scaling}\label{Sec:FShS}
Inspired by the results that we obtained for the Lebwohl-Lasher 
model~\cite{FarinasParedesBerche03} or the XY 
model~\cite{BercheFarinasParedes02,Berche02,BercheShchur04}, we will now 
produce a 
complementary study
using a rescaling of the density profiles. For that purpose, we 
assume  the existence of a critical 
phase at low temperatures as suggested by the temperature dependence
and FSS results. If the assumption is revealed incorrect, 
we will be led to some
inconsistency. 

The existence of a scale-invariant 
low temperature critical phase
leads to conformally covariant density profiles or correlation 
functions at {\em any temperature below the 
transition $T_{c}$}.
It is then advantageous to deduce the functional expression of the 
correlation functions or density profiles in a restricted geometry 
adapted to numerical simulations from a conformal mapping $w(z)$:
\begin{equation}
G(w_1,w_2) =
| w'(z_{1})|^{-x_{\sigma}} | w'(z_{2})|^{-x_{\sigma}}
G(z_1,z_2)
\end{equation}
Here, $w$ labels the lattice sites in the transformed geometry (the 
one where the computations are really performed), $z$ 
is the corresponding point in the original one (usually the 
infinite plane where  the two-point correlations
 take the standard power-law 
expression $G(z_1,z_2)
\sim |z_1 -z_2 |^{-\eta_{\sigma}}$), and $x_{\sigma} = \frac{1}{2}\eta_{\sigma}$
 is the scaling dimension associated to the scaling field under consideration.
The interest of such an approach lies in the full inclusion  
of the changes due to shape effects in the 
functional expression and we may
copy the terminology FSS and call it {\em Finite-Shape Scaling}. 
%%% figure %%%
\begin{figure}[ht]
  \centering
  \begin{minipage}{\textwidth}
  \qquad \epsfig{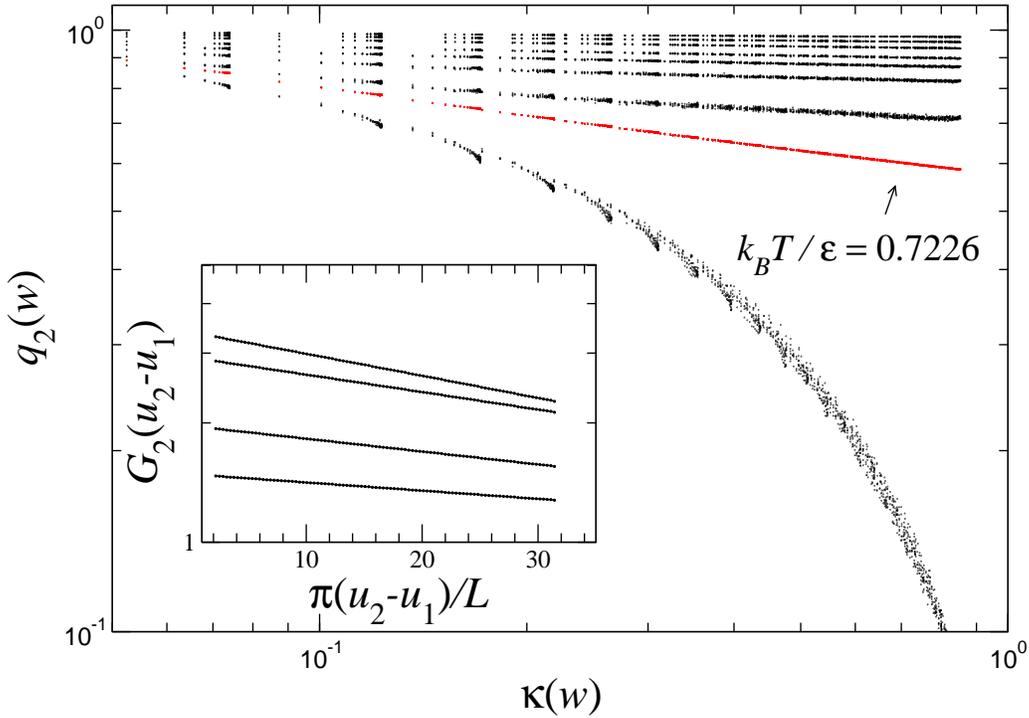}
  \end{minipage}
  \caption{Finite-Shape Scaling behaviour ($L=64$) of the order parameter 
	density profile $q_2(w)$  
	for various 
	temperatures $k_BT/\epsilon$ below the transition temperature 
	and one temperature above.
	The transition takes place around the value 0.70.
	The insert shows (on a semi-log scale) 
	the exponential decay of correlation function (below the transition)
	along a torus as explained in the text. 
	}
	\label{Fig5}
\end{figure}
%%% end figure %%%
Rather than two-point correlation functions, it 
is even more convenient to work with density 
profiles $m(w)$ in a finite system with symmetry breaking fields along some 
surfaces in order to induce a non-vanishing local order parameter in 
the bulk. 
The density $m(w)$ will be
$q_2({\bf r})=\<Q_2(\bsigma_\r\cdot{\bf h}_{\partial\Lambda})\>$ 
or higher rank nematisation 
$q_4({\bf r})=\<Q_4(\bsigma_\r\cdot{\bf h}_{\partial\Lambda})\>$.
In the case of a square lattice $\Lambda$ of size 
$L \times L$, with fixed boundary conditions along the four 
edges $\partial \Lambda$, one expects
\begin{eqnarray}
	m(w)& \sim&
 	[\kappa(w)]^{-\frac{1}{2} \eta_{\sigma}}
	\nonumber\\
	\kappa(w)  &=&  
	\Im{\rm m} \left[ {\rm sn} \frac{2{\rm K}w}{L} \right]  \times \left| 
	\left( 1 - {\rm sn}^{2} 
	\frac{2{\rm K}w}{L} \right) 
	\left(1 - k^{2} {\rm sn}^{2} 
	\frac{2{\rm K}w}{L} \right) \right|^{-\frac{1}{2}}
	\label{eq:squaremapping}
\end{eqnarray}
where $w$ stands for lattice site $\r$.
This expression easily follows from the expression of the order parameter 
profile decaying in the upper half-plane from a distant surface of 
spins constantly fixed in a given direction, 
$m(z) \sim y^{-x_{\sigma}}$, and from the conformal 
transformation of the upper 
half-plane
$z=x+iy$, $(0 \leq y < \infty)$, inside a square 
$w=u+iv$ of size 
$L \times L$, $ (-L/2 \leq u \leq L/2, 0 \leq v \leq L)$,
with open boundary conditions along the four edges, realized by a 
Schwarz-Christoffel transformation 
\begin{equation}
w(z) = \frac{L}{2{\rm K}} {\rm F}(z,k), \;  \; \; \; \; \;  z={\rm sn} 
\left( \frac{2{\rm K}w}{L} \right).
\end{equation}
Here, F$(z, k)$ is the elliptic integral of the first kind, 
${\rm sn}(2{\rm K}w/L)$ the Jacobian elliptic sine, K = K$(k) = $F$(1, k)$
 the complete elliptic 
integral of the first kind, and the modulus $k=0.171573$ 
depends on the aspect ratio of 
$\Lambda$ (here $1$).

The procedure is now to fit numerical data of the order parameter profile 
against expression (\ref{eq:squaremapping}).
The result 
for the density profile
of $q_2(w)$ is shown on a log-log scale in Fig.~\ref{Fig5}. 
At low temperatures, the resulting straight lines confirm the existence of
a critical phase (a linear behaviour on this scale results from an algebraic
decay in the original semi-infinite geometry). 
Above the deconfining transition (assuming that it 
is indeed the driving mechanism 
of the transition), the decay becomes faster, 
indicating a paramagnetic phase. 
The transition is 
approximately located at a temperature $k_BT_c/\epsilon\simeq 0.70-0.75$.
The scenario is eventually completely consistent with a BKT transition.
Furthermore the $\eta$ exponent again follows from the scaling of the
density profile in Eq.~(\ref{eq:squaremapping}) which provides an alternate 
determination of this quantity. 

Another famous conformal mapping which has been applied to many 
two-dimensional critical systems is the logarithmic transformation
$w(z)=\frac {L}{2\pi}\ln z=\frac {L}{2\pi}\ln \rho+i\frac{L\varphi}{2\pi}$.
It maps the infinite plane onto an infinitely long cylinder of perimeter $L$,
and due to the one-dimensional character of this latter geometry, the 
correlation functions along the axis of the cylinder (let say in terms of
the variable $u=\frac {L}{2\pi}\ln \rho$)
decay exponentially 
at criticality, $G(u_1,u_2)\sim\exp[-(u_2-u_1)/\xi]$. The interesting
result which makes this technique powerful is that the correlation length 
amplitude on the strip is universal and only determined by the
corresponding $\eta$ exponent, $\xi=\frac{L}{\pi\eta}$.
This relation is known as the gap-exponent relation in the context of
quantum chains in $1+1$ dimensions. It was conjectured by several 
authors~\cite{PichardSarma,DerridaDeSeze,Luck} before Cardy proved 
it~\cite{Cardy84}. 
Using MC simulations we cannot of course produce an infinitely long cylinder.
It is however possible to perform simulations inside a rectangle $L_1\times
L$ with $L_1\gg L$ and periodic boundary conditions in both space 
directions (we get a very long torus). Due to the exponential decay of
the correlation length, it is not necessary to explore really long
distances $u_2-u_1$ (typically $u_2-u_1\le 10L$). A finite
torus of long perimeter $10^3L$ thus only produces insignificant
finite-size corrections to the gap-exponent relation. We performed
the simulations at different temperatures in a system of size 
$10\times 10000$ and extracted the correlation function exponent from
the linear behaviour
\be\ln\<Q_2[\cos(\theta_{u_2}-\theta_{u_1}]\>={\rm const}-\frac{\pi
	\eta_{q_2}}{L}(u_2-u_1).\ee
This linear behaviour (in terms of the variable $u_2-u_1$) is shown in the
insert of Fig.~\ref{Fig5} where for simplicity 
$\<Q_2[\cos(\theta_{u_2}-\theta_{u_1}]\>$ is denoted $G_2(u_2-u_1)$.
It is impossible to apply this technique up to the transition temperature, 
since the strip system contains quite a large number of spins 
($10^5$ while simulations in the square geometry are performed up to 
typically $10^4$ spins) and the autocorrelation time increases too fast.

\section{Behaviour at the deconfining transition}
\begin{figure}[ht]
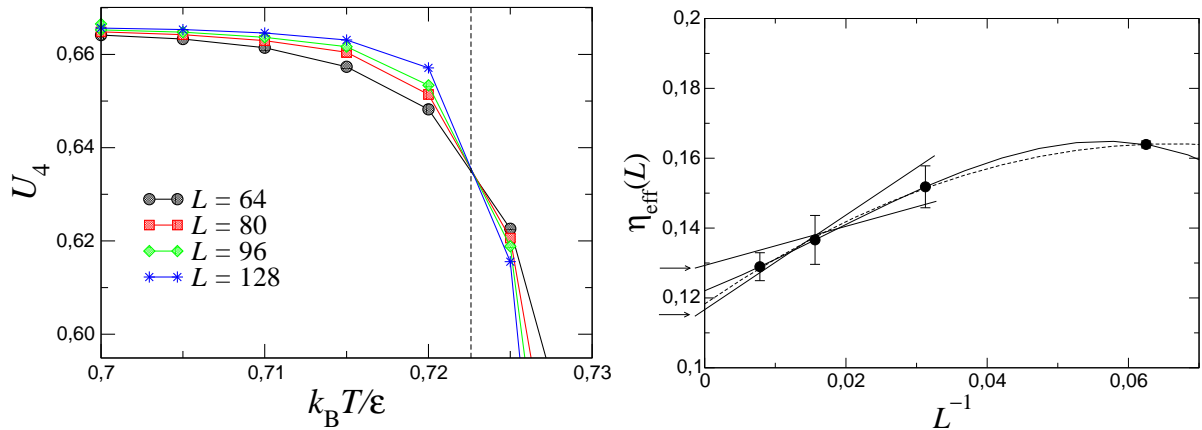

  \centering
  \begin{minipage}{\textwidth}
  \epsfig{file=Fig_U4.eps,width=0.5\textwidth}
  \epsfig{file=Fig_Eta-T_BKT.eps,width=0.48\textwidth}
  \end{minipage}
  \caption{Left: Crossing of the Binder parameter at the deconfining transition
	at a temperature
	$k_BT_{\rm BKT}/\epsilon=0.7226$. Right: Size dependence of the
	correlation function exponent at $T_{\rm BKT}$ ($L=16$, 32, 64, 128)
	and extrapolation
	to the thermodynamic limit.}
	\label{Fig8}
\end{figure}

Not only the low temperature behaviour is interesting. The value of the
$\eta$ exponent at the BKT transition where some deconfining 
mechanism should lead to the proliferation of unbinded topological defects is
also of primary interest. For that purpose, an accurate value of the
transition temperature is needed.
We performed a study of the crossing of $U_4$ Binder cumulant for very
large statistics ($30\times 10^6$ MCS) and large system sizes (squares
of $L=64$, 80, 96 and 128 with periodic boundary conditions). The results 
shown in Fig.~\ref{Fig8} indicate a transition temperature of
$k_BT_{\rm BKT}/\epsilon=0.7226$.

%%% figure %%%
%%% end figure %%%

Then this temperature is used to perform Finite-Shape Scaling using
the technique already employed for the investigation of the low
temperature phase, namely the algebraic decay of density profiles 
inside a square with fixed bounday conditions. 
These new simulations are really time-consuming, since the autocorrelation
time increases in the low temperature phase when $T$ evolves towards
the deconfining transition and a rather large number of Monte Carlo steps
is needed to get a satisfying number of independent measurements.
For sizes $L=16$, 32, 64, we used $10^6$ MCS for thermalization and
$30.10^6$ for measurements, while ``only'' $20.10^6$ for the largest size 128.
For technical reasons, the exponential
decay of two-point correlation functions along the torus cannot be applied
at the BKT transition, 
since as already mentioned the system size being quite larger than in a square
geometry, the number of MC
iterations required is by far too large.
In Fig.~\ref{Fig8} we plot the ``effective'' exponent $\eta_{\rm eff}(L)$
measured at $T_{\rm BKT}$ for different system sizes as a function of
the inverse size. An estimate of the thermodynamic limit value ($L\to\infty$)
can be made using a polynomial fit (the results of  quadratic
and cubic fits are respectively $0.118$ and
$0.122$). It is safer to keep the three largest sizes available,
$L=32$, 64, and 128, for which a linear dependence of $\eta_{\rm eff}(L)$
with $L^{-1}$ is observed. 
Taking into account the error bars, crossing the
extreme straight lines leads to 
the following value for the correlation function exponent at the
deconfining transition
\be\eta_{q_2}(T_{\rm BKT})= 0.122\pm 0.007.\ee
This value is essentially half the Kosterlitz value for the $XY$ model.

\section{Conclusion}\label{sec:ccl}
\begin{figure}[ht]
  \centering
  \begin{minipage}{\textwidth}
  \epsfig{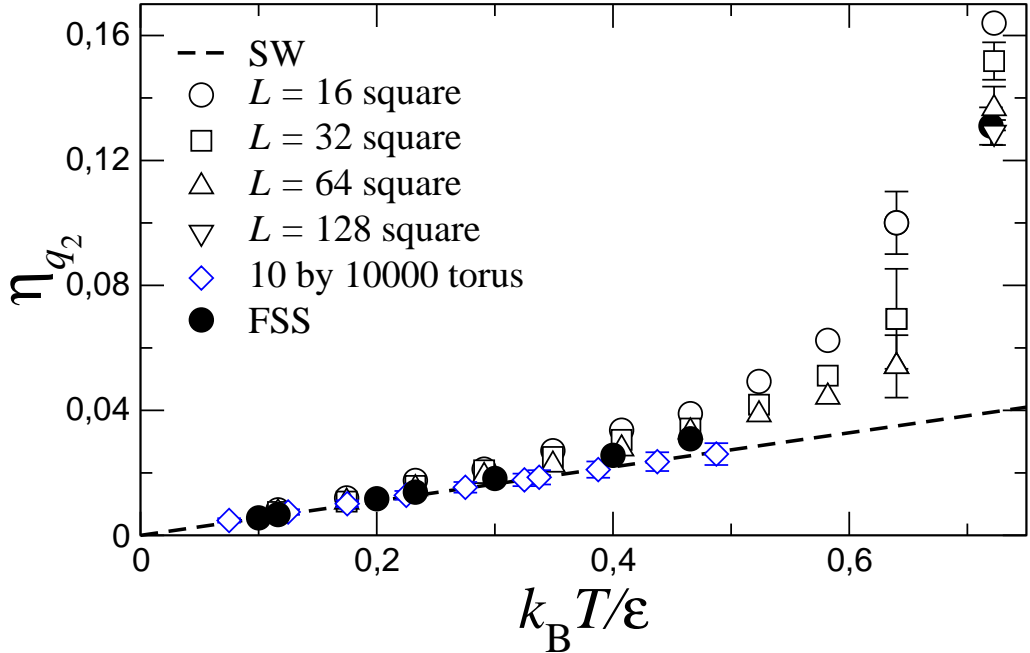}
  \end{minipage}
  \caption{Temperature variation of the correlation function exponent  
	$\eta_{q_2}(T)$  deduced from 
	conformal rescaling (open symbols) and FSS (filled symbols).
	Open triangles and diamonds, which correspond the largest systems
	where we appplied FShS seem quite reliable.  
	The dashed lines shows the
	result of the spin-wave approximation
	$\eta_{q_2}(T)=\eta^{(2)}_{2\frac{128}{11}}=\frac{11}{64}
	\frac{k_BT}{\pi\epsilon}$.
	}
	\label{Fig6and7}
\end{figure}

In Fig.~\ref{Fig6and7}  
we plot as a function
of the temperature the exponent $\eta_{q_2}(T)$
measured after conformal
rescaling of the density profile and correlation functions
at different system sizes and the FSS
determination which follows from Fig.~\ref{Fig3}. 
Together with the exponent determined numerically, we report the result 
of the spin-wave approximation, shown in dotted
line. The larger the size of the system, the better the agreement. 
Similar results (not shown here) are
measured in the case of the higher-order nematisation, $q_4(T)$.

%%% figure %%%
%%% end figure %%%

According to these results, the main outcome of the present 
work is the following:
\begin{description}
\item{-} $P_4\ O(2)$ model displays a BKT-like transition with QLRO 
in the LT phase
where SWA nicely fits the nematisation temperature-dependent 
exponent $\eta_{q_2}(T)$
when $T\to 0$.
\end{description}
The LT phase of the model is thus disordered. Nevertheless, in what we
will call a {\em physical limit}, i.e. a finite but quite large 
size where the system
contains a macroscopic number of spins, it is useful to consider 
that the system is
partially ordered, as Fig.~\ref{Fig3bis} seems to indicate. 

These results find a partial interpretation through a naive comparison 
with clock model  
in 2D. Increasing the order of the interaction polynomial indeed 
increases the number
of deep wells which stabilise the relative orientation of neighbouring spins. 
One is thus led to a system 
which is 
quite similar to
a planar clock model with a finite number of states, unless the fact 
that here we keep a 
continuous spin symmetry which prevents from any ``magnetic'' 
long-range order at 
finite temperature.
The clock model is known to be in the Potts universality class when 
$q=3$, but at 
$q\ge 4$, it displays a QLRO phase before conventional ordering at 
lower temperatures~\cite{JoseEtAl77}.
Combining these results  with the requirements of Mermin-Wagner-Hohenberg
theorem in the case of continuous spin symmetry gives a natural
framework for the comprehension of our results for $2-$component spin systems.
Whatever the nearest-neighbour interaction (in $P_1$, $P_2$ or
$P_4$) their behaviour seems to be always described by a BKT transition.
The similar observation that a two-component nematic model renormalises in two
dimensions towards the $XY$ model was already reported in 
Ref.~\cite{NelsonPelcovits77}.
The transition is likely driven by a mechanism of condensation of defects,
like in the $XY$ model, but due to the local $Z_2$ symmetry not only
usual vortices carrying a charge $\pm 1$ are stable, but also disclination
points carrying charges $\pm 1/2$ should be stable. The role of these defects
might be studied in a similar way than in the recent work of
Dutta and Roy~\cite{DuttaRoy04}, by the comparison of the the transition
in the pure model and in a modified version where a chemical potential
is artificially introduced in order to control the presence of defects.

\section*{Acknowledgement} The work of A.I.F.S. 
is supported by a PCP cooperation
programme (``{\em Fluides p\'etroliers}'') between France and 
Venezuela. 
Thanks to the support from CINES in Montpellier for computational time. 
We benefited from instructive correspondence on $O(n)$
models with H. Kawamura, 
A. Pelissetto, 
D. Mouhanna and S. Korshunov who are gratefully acknowledged. 
B.B. is also indebted to Yu. Holovatch for stimulating  discussions 
at the occasion 
of one of his visits in Nancy.

%%%%%%%%%%%%%%%%%%%%%%%%%%%%%%%%%%%%%%%%%%%%%%%%%%%%%%%%%%%
%        REFERENCES
%%%%%%%%%%%%%%%%%%%%%%%%%%%%%%%%%%%%%%%%%%%%%%%%%%%%%%%%%%

\end{document}